\documentclass[amsmath,amssymb, reprint]{revtex4-1}

\usepackage{graphicx}% Include figure files
\usepackage{dcolumn}% Align table columns on decimal point
\usepackage{bm}% bold math
\usepackage[utf8]{inputenc}
\usepackage[T1]{fontenc}
\usepackage{mathptmx}
\usepackage{etoolbox}
\usepackage{amsmath}
\usepackage{braket}
\makeatletter
\def\@email#1#2{%
 \endgroup
 \patchcmd{\titleblock@produce}
  {\frontmatter@RRAPformat}
  {\frontmatter@RRAPformat{\produce@RRAP{*#1\href{mailto:#2}{#2}}}\frontmatter@RRAPformat}
  {}{}
}%
\makeatother
\begin{document}

%\preprint{AIP/123-QED}

\title[]{A Comment on\\ "Algebraic approach to the Tavis-Cummings model with three modes of oscillation" [J. Math. Phys. 59, 073506 (2018)]}
% Force line breaks with \\
\author{Viani S. Morales-Guzman}
 \email{viani.morales@correo.nucleares.unam.mx}
\author{Jorge G. Hirsch}%
\email{hirsch@nucleares.unam.mx}
 \affiliation{Instituto de Ciencias Nucleares, Universidad Nacional Autónoma de México}

\date{\today}% It is always \today, today,

\begin{abstract}
Choreño et al. [J. Math. Phys. 59, 073506 (2018)] reported analytic solutions to the resonant case of the Tavis-Cummings model, obtained by mapping it to a Hamiltonian with three bosons and applying a Bogoliubov transformation. This comment points out that the Bogoliubov transformation employed is not unitary, cannot be inverted, and cannot enforce the symmetries of the model.
\end{abstract}

\maketitle

\iffalse
\begin{quotation}
The ``lead paragraph'' is encapsulated with the \LaTeX\ 
\verb+quotation+ environment and is formatted as a single paragraph before the first section heading. 
(The \verb+quotation+ environment reverts to its usual meaning after the first sectioning command.) 
Note that numbered references are allowed in the lead paragraph.
%
The lead paragraph will only be found in an article being prepared for the journal \textit{Chaos}.
\end{quotation}
\fi

\section{\label{sec:intro}Introduction} 
In contrast with the known approximated analytic solutions to the Tavis-Cummings (TC) model \cite{Garraway2011}; in the article "Algebraic approach to the Tavis-Cummings model with three modes of oscillation", Choreño et al.\cite{Choreno-2018} present, for the first time, exact analytical expressions for the eigenergies and the eigenfunctions. 
Employing the Schwinger representation of angular momentum operators in terms of boson operators, the TC Hamiltonian is mapped to a Hamiltonian with three bosons. Three exact solutions are obtained employing a Bogoliubov transformation, normal-mode operators and tilting transformation.  

In what follows, it is shown that in the three cases the transformation employed are not unitary, cannot be inverted, and cannot be used to obtain meaningful solutions associated with the symmetries of the TC Hamiltonian. 

The Tavis-Cummings Hamiltonian, at resonance, with $\hbar=1$, reads
\begin{equation}
    H_{TC}=\omega \hat{c}^\dagger \hat{c} + \omega \hat{J}_z + \kappa (\hat{c} \hat{J}_+ + \hat{c}^\dagger \hat{J}_-).
\end{equation}
It can be mapped into the three boson Hamiltonian 
\begin{equation}
    H= \omega \hat{a}^\dagger \hat{a} +\omega \hat{b}^\dagger \hat{b}+ \omega \hat{c}^\dagger \hat{c} + g(\hat{a}^\dagger \hat{b}\hat{c} +\hat{a} \hat{b}^\dagger\hat{c}^\dagger), 
    \label{eq:TW}
\end{equation}
employing the Schwinger representation 
\begin{equation}
\hat{J}_+=\hat{a}^\dagger \hat{b},  \,\, \hat{J}_-=\hat{a} \hat{b}^\dagger, \,\, \hat{J}_z= \frac 1 2 \left(\hat{a}^\dagger \hat{a} - \hat{b}^\dagger \hat{b}\right) 
\end{equation}
In the above equation
$g = \kappa$ is the coupling constant, and $\hat{a}$, $\hat{a}^\dagger$, $\hat{b}$, $\hat{b}^\dagger$, $\hat{c}$, $\hat{c}^\dagger$ are bosonic annihilation and creation operators.
 Additionaly, Hamiltonian (1) commutes with the operator $\hat\Lambda \equiv \hat{c}^\dagger \hat{c} + J_z$, with eigenvalues $\lambda$.

There are two constrictions that the number of bosons must satisfy: 
\begin{equation}
n_a + n_b = 2j,
 \,\, \,\,
n_c + \frac 1 2 (n_a - nb) =  \lambda .
 \end{equation}
The values of $j$ and $\lambda$ determine independent subspaces of the Hilbert space.

In section III. A., Choreño et al. introduce a Bogoliubov transformation 
\begin{equation}
\begin{split}
    \hat{b}&= \hat{f} \cosh{r}+ \hat{d}^\dagger e^{-i\theta} \sinh{r}\\    \hat{c}&=\hat{d} \cosh{r} + \hat{f}^\dagger e^{-i \theta} \sinh{r},
\end{split},
\label{eq:BTtrans}.
\end{equation}
The transformed Hamiltonian presented in the article is, 
\begin{equation}
\begin{split}
        H^\prime=& \left[\omega(1+2 \sinh^2{r})+ \frac{g}{2}(\hat{a}e^{i \theta}+\hat{a}^\dagger e^{-i \theta})\sinh{2r}\right]\\&(\hat{f}^\dagger\hat{f}+\hat{d}^\dagger\hat{d}+1)\\
        &\left[e^{i \theta} \sinh{2r} + g^*\hat{a}^\dagger \cosh^2{r}+ g\hat{a}e^{2 i \theta} \sinh^2{r}\right] \hat{f} \hat{d}\\
        &\left[e^{-i \theta} \sinh{2r} + g\hat{a} \cosh^2{r}+ g^*\hat{a}^\dagger e^{-2 i \theta} \sinh^2{r}\right] \hat{f}^\dagger \hat{d}^\dagger\\
        & \omega_1 \hat{a}^\dagger\hat{a} -\omega
    \label{eq:Hamprima}
\end{split}
\end{equation}

The parameters $\hat{r}$ and $\hat{\theta}$ are selected to cancel the terms multiplying $\hat{f} \hat{d}$ and $\hat{f}^\dagger \hat{d}^\dagger$ in
\eqref{eq:Hamprima}. They can be written in terms of $\hat{u}$ and $\hat{v}$ as 
\begin{equation}
\begin{split}
    \hat{u}&\equiv \cosh{\hat{r}}=\frac{\omega}{\sqrt{\omega^2-|g|^2\hat{a}^\dagger \hat{a}}}\\
    \hat{v}& \equiv e^{-i \hat{\theta}}\sinh{\hat{r}}=\sqrt{\frac{\hat{a}}{\hat{a}^\dagger}} \frac{|g|^2\sqrt{\hat{a}^\dagger \hat{a}}}{\sqrt{\omega^2-|g|^2\hat{a}^\dagger \hat{a}}}.
\end{split}
\label{eq:r,theta}
\end{equation}

Using \eqref{eq:r,theta} in \eqref{eq:Hamprima}, the Hamiltonian takes the diagonal form
\begin{equation}
    H^\prime= \omega_1 \hat{a}^\dagger \hat{a} + \sqrt{\omega^2-g^2\hat{a}^\dagger \hat{a}}(\hat{f}^\dagger\hat{f}+\hat{d}^\dagger\hat{d}+1 )- \omega.
    \label{eq:HBT}
\end{equation}
with analytical eigenvalues and eigenstates
\begin{align}
    E^\prime=& \sqrt{\omega^2-g^2n_a}(n_f+n_d+1)+ \omega_1 n_a -\omega \label{eq:E_BT}\\
    \Psi^\prime=&\psi_{n_a}(x) \otimes \psi_{n_l, m_n}(\rho, \phi);
    \label{eq:phi}
\end{align}
where $\psi_{n_a}(x)$ are the eigenfunctions of the one-dimensional harmonic oscillator and $\psi_{n_l, m_n}(\rho, \phi)$ are the eigenfunctions of the 2D harmonic oscillator. 

The problem with the above deduction is that the transformation (5) is not unitary. It implies that the new operators $\hat{d},\hat{f}$ do not satisfy bosonic commutation relations.

To probe this point, observe that the transformation coefficients $\hat{u},\hat{v}$ in (5) are operators, but in Ref\cite{Choreno-2018} where treated as scalars.
Given that $\hat{a}$ and $\hat{a}^\dagger$ do not commute with $\hat{n}_a$, the determinant of the proposed transformation (5) is
\begin{align}
    \hat{u} \hat{u}^\dagger - \hat{v} \hat{v}^\dagger&=1\\&-|g|^2\left[~\sqrt{\frac{\hat{a}}{\hat{a}^\dagger}}, \sqrt{\frac{\hat{n_a}}{\omega^2-|g|^2\hat{n_a}}}~\right] \sqrt{\frac{\hat{a}^\dagger}{\hat{a}}}\sqrt{\frac{\hat{n_a}}{\omega^2-|g|^2\hat{n_a}}}\nonumber \\&\neq 1.
\end{align}
therefore, \textbf{the transformation \eqref{eq:BTtrans} is non-unitary}. 

As a consequence, following \eqref{eq:BTtrans}, $\hat{b}= \hat{u}\hat{f}+ \hat{v}\hat{d}^\dagger$, the commutator for $\hat{b}$ and $ \hat{b}^\dagger$ is
\begin{equation}
\begin{split}
    [\hat{b}, \hat{b}^\dagger]&=\hat{u} \hat{u}^\dagger [\hat{f}, \hat{f}^\dagger]+\hat{u} \hat{v}^\dagger[\hat{f}, \hat{d}]+\hat{v} \hat{u}^\dagger[\hat{d}^\dagger, \hat{f}^\dagger]+\hat{v} \hat{v}^\dagger [\hat{d}^\dagger, \hat{d}]\\&=\hat{u} \hat{u}^\dagger [\hat{f}, \hat{f}^\dagger] - \hat{v} \hat{v}^\dagger[\hat{d}, \hat{d}^\dagger].
\end{split}    \label{eq:conmutadorbb}
\end{equation}
From \eqref{eq:conmutadorbb} it is noted that \textbf{if $\hat{f}$, $\hat{f}^\dagger$, $\hat{d}$ and $\hat{d}^\dagger$ are bosonic operators, the operators $\hat{b}$ and $\hat{b}^\dagger$  written in terms of the transformation do not satisfy the bosonic algebra  and viceversa}; the same happens for $\hat{c}$ and $\hat{c}^\dagger$. 

It could still be interesting to analyze if the eigenenergies, given in Eq. (9) as simple functions of the number of bosons $n_a, n_d$ and $n_f$, can be comparared with the exact energies obtained by numerical diagonalization. But this task is impossible, because, given that transformation (5) is not invertible, there is no way to map  $n_d, n_f$ to the number of original bosons $n_b, n_c$. It follows that it is not possible to select the subspaces of the Hilbert space, with fixed values of $j$ and $\lambda$, associated with the numbers $n_d, n_f$.  

It is relevant to mention that the same problem occurs for the other two transformations presented in Ref\cite{Choreno-2018}. In the tilting transformation (21) the coherent state parameters $\hat{\theta}$ and $\hat{\phi}$ do not commute, and the normal mode operators defined in (34) do not commute, because according to (39) and (41), the ''constant'' $X$ is an operator $\hat{X}$ which does not commute with $\hat{X}^\dagger$.

It is worth to point out that, in another article by the same authors, entitled "Matrix diagonalization and exact solution of the k-photon Jaynes-Cummings model" \cite{Choreno-2018-JC}, a similar procedure is used to solve the k-photon Jaynes-Cummings Hamiltonian. But in this case the correct eigensystem is obtained.

Choreño et al. start by using the interaction Hamiltonian for the model\footnote{We refer to the two-photon Jaynes-Cummings model for simplicity, the generalization is analogous.},
\begin{equation}
    H_I=\hbar(\frac{\omega_0}{2} - \omega) \sigma_z+g (\sigma_+ (\hat{a})^2 + \sigma_- (\hat{a}^\dagger)^2 ), 
    \label{eq:HI}
\end{equation}
 where $\sigma_{z,\pm}$ are the Pauli matrices, $\omega_0$ and $\omega$ are the transition and field frequency respectively. A transformation is applied, 
\begin{equation}
    H_{I}^\prime= D^\dagger(\xi) H_I D(\xi),
    \label{eq:HItrans}
\end{equation}
with 
\begin{equation}
    D(\xi)= exp(\xi \sigma_+-\xi^*\sigma_-),
    \label{eq:displace}
\end{equation}
where $\xi=-\frac{1}{2} \hat{r} e^{-i \hat{\theta}}$. 

After expressing the Hamiltonian in matrix form, the non-diagonal terms are set to zero and the equations are solved for $\hat{r}$ and $\hat{\theta}$ which result dependent on $\hat{a}$ y $\hat{a}^\dagger$. By substituting these parameters in the transformed Schrödinger equation,

\begin{equation}
    D^\dagger(\xi) H_I D(\xi)D^\dagger(\xi) \Psi = E_I D^\dagger(\xi) \Psi
\end{equation}
one can obtain the eigensystem accurately. 

Why is it that the same procedure as in Tavis-Cummings led them to the correct eigensystem? Because in this case, \textbf{the transformation used to solve the Jaynes-Cummings model is unitary}. In its matrix form, 
\begin{equation}
    D(\xi)= \begin{pmatrix}
        \frac{1}{\sqrt{2}}\sqrt{1+ \frac{\Delta}{E_I}} & -\frac{1}{\sqrt{2}}\sqrt{1- \frac{\Delta}{E_I}}\frac{\sqrt{(n+1)(n+2)}}{{(\hat{a}^\dagger)^2}}\\  \frac{1}{\sqrt{2}}\sqrt{1- \frac{\Delta}{E_I}}\frac{(\hat{a}^\dagger)^2}{\sqrt{(n+1)(n+2)}} &\frac{1}{\sqrt{2}}\sqrt{1+ \frac{\Delta}{E_I}}
    \end{pmatrix}.
    \label{eq:Dxi}
\end{equation}
Expressing it in the appropriate basis,
\begin{equation}
\{\ket{\psi_{1n}} \equiv  \ket{n,g}, \{\ket{\psi_{2m}} \equiv\ket{m,e}\},
\label{eq:basis}
\end{equation} 
$D(\xi)$ takes the form, 
\begin{equation}
    D(\xi)=\begin{pmatrix}
       \frac{1}{\sqrt{2}}\sqrt{1+ \frac{\Delta}{E_I}} & -\frac{1}{\sqrt{2}}\sqrt{1- \frac{\Delta}{E_I}}\\  \frac{1}{\sqrt{2}}\sqrt{1- \frac{\Delta}{E_I}} &\frac{1}{\sqrt{2}}\sqrt{1+ \frac{\Delta}{E_I}}
    \end{pmatrix},
\end{equation}
which is clearly unitary. 

Although the procedure presented above gives a valid solution, it is useful to remind the reader that the two-photon (k-photon) Jaynes-Cummings Hamiltonian is always a $2 \times 2$ matrix which can be easily diagonalized, expressing the Hamiltonian in the basis \eqref{eq:basis},

\begin{equation}
    H_I=\begin{pmatrix}
\hbar(\frac{\omega_0}{2} - \omega) & g \sqrt{(m)(m-1)} \delta_{n, m-2} \\
g \sqrt{(n+1)(n+2)} \delta_{n+2, m} & -\hbar(\frac{\omega_0}{2} - \omega)
\end{pmatrix}
\end{equation}
to obtain the exact eigenvalues
\begin{equation}
    E_{\pm}(n)= \hbar \omega (n+1) \pm\sqrt{\Omega^2+g^2(n+1)(n+2)},
\end{equation}
with $n=0,1,2,...$. 

In conclusion, the Bogoliubov transformations employed to solve the Jaynes-Cummings model provides correct results, and can be seen as a cumbersome procedure to diagonalize a $2 \times 2$ matrix. Instead, when the same formalism is generalized for the Tavis-Cummings model, the transformations are not unitary and the ''analytic'' results are not valid.

\section*{conflict of interest}
The authors have no conflicts of interest to disclose.

\section*{Author's Contributions}
VSMG made the formal analysis and the original draft writing, JGH contributed to the conceptualization, writing and editing.

\begin{acknowledgments}
We acknowledge partial financial support from DGAPA-UNAM project PAPIIT IN109523.
\end{acknowledgments}

\bibliographystyle{ieeetr}
%\bibliography{BIBLIOGRAPHY}
\providecommand{\noopsort}[1]{}\providecommand{\singleletter}[1]{#1}%

\end{document}